# Numerical modeling of roll structures in mesoscale vortexes over the Black Sea

## Abstract

This paper is a case study of horizontal atmospheric rolls that formed over the Black Sea on 16 August 2007. The rolls were discovered in WRF modeling results for a mesoscale cyclone that originated over the sea on 15 August 2007. The roll formation mechanisms, such as Rayleigh-Benard convective instability, dynamic instability, advection and stretching of vertical velocity field inhomogeneities, are considered.

It is shown that indeed convective instability played an important role in the roll formation but dynamic instability did not occur. In order to distinguish other possible mechanisms of the roll formation numerical experiments were performed. In these experiments sea surface temperature in the initial conditions was decreased in order to prevent convective instability. Even though convective instability was suppressed roll-like structures still appeared in the modeling results, although their height and circulation velocity were smaller than in the control run. It was found that these structures were caused by advection and stretching of vertical velocity field inhomogeneities (i.e. small areas of strong updrafts or downdrafts).

## Introduction

At the present time atmospheric mesoscale circulation in the Black Sea Region is not properly studied mainly because observational and re-analysis atmospheric data with high spatial resolution were unavailable. There are only a few studies related to the Black Sea mesoscale cyclones [1 – 4]. In [1] statistical analysis of mesoscale cyclones over the Black Sea was performed using a dataset on surface wind fields for the Black Sea Region for the period of 1958-1996 with 25 km-spatial and 1 hour-temporal resolution. The dataset was obtained using regional climate model PRECIS. The model was initialized with the ERA-40 global re-analysis data. In [1] it was found that there are two main types of mesoscale cyclones in the Black Sea Region, which form over the sea near the Crimean coast and near the Caucasian coast (Figure 1). According to their place of origin these mesoscale cyclones were called Caucasian and Crimean respectively. In [2] the authors performed a detailed case study of a Crimean mesoscale cyclone detected in numerical modeling results on 15 August 2007. Circulation rolls that formed in the planetary boundary layer (PBL) of the cyclone required additional investigation. This paper is a sequel to [2].

Long and narrow circulation rolls are often detected in atmospheric modeling results for the Black Sea Region. The roll parameters are as follows: width is 3-5 km, length is up to 50 km, height is 700-1000 m, vertical velocity of air ascend and subsidence is ~±0.2 $m \cdot s^{-1}$. Rolls can be parallel to the wind direction or make an angle with it.

Usually rolls form as a result of convective and dynamic instability [5]. In the former case rolls develop by consuming convective available potential energy (CAPE) (i.e. due to buoyancy force work), and in the latter case − by consuming Reynolds-averaged flow kinetic energy. As usual, both mechanisms act simultaneously and it is difficult to distinguish them. According to [5] both mechanisms can play equal roles in the roll development.

**Numerical model**

The main investigation method is numerical modeling because observational data with high spatial resolution is unavailable. The simulation is performed using a numerical model of regional atmospheric circulation ARW (Advanced Research WRF) 3.3.1 (Skamarock *et al.*, 2008) with three nested domains. The domains have horizontal resolutions of 9, 3, and 1 km. There are 37 irregularly spaced η-levels in the vertical, with a higher resolution in the lower troposphere.

The following parametrization schemes are used: the RRTM (Rapid Radiative Transfer Model) and Dudhia schemes of longwave and shortwave radiation transfer respectively; Single-moment 3-class microphysics scheme; the Kain-Fritsch cumulus parametrization scheme for the domains with 9 and 3 km resolution (Skamarock *et al.*, 2008) (there is no need for cumulus parametrization in the domain with 1 km-resolution).

The PBL is parametrized using the Yonsei university scheme (Hong *et al.*, 2006). A profile function of a vertical momentum diffusion coefficient, $K_z$, in this scheme is proportional to $z(1 - z/H)^2$, where H is PBL thickness. Non-local mixing processes are accounted for by countergradient terms proportional to a flow from the surface. Coefficients of heat and moist diffusion are calculated using $K_z$ and the variable Prandtl number. Non-dimensional wind, heat, and moist profile functions are defined by the well-known Monin-Obukhov relations. Entrainment processes at the PBL top are also taken into account. In the free atmosphere a closure scheme, in which vertical diffusion coefficients are proportional to mixing length squared, vertical gradient of velocity, and the local Richardson number, is used. In the entrainment zone a diffusion coefficient is defined as the geometric mean of diffusion coefficients in the free atmosphere and at the PBL top. The PBL height is defined by the critical bulk Richardson number.

A horizontal momentum diffusion coefficient, $K_h$, is defined as $K_h = C^2 \cdot \Delta x \cdot \Delta y \cdot \sigma$, where C is a constant with a typical value of 0,25; Δx, Δy is horizontal spatial resolution; σ is horizontal wind field deformation.

Initial conditions for all three domains and boundary conditions for the outermost domain are chosen from the FNL (Final) Operational Global analysis with a spatial resolution of 0.5º. After relaxation of the initial conditions the evolution of atmospheric processes in the domains is governed only by the lateral conditions, which are updated every 6 hours for the outermost domain.

**Circulation rolls in the control run**

Let us consider the Crimean vortex that formed on 15 August 2007 over the Black Sea. The vortex originated at night near the south part of the Crimean peninsula, then moved away from the shore and gradually dissipating traveled a distance of 200 − 250 km over the sea. The vortex structure and evolution was studied in detail in [2]. In this paper we only study the circulation rolls that formed in the vortex PBL. We shall call the modeling results used in [2] a control run.

Fig.2 shows vertical and horizontal wind fields at height of 400 m where vertical velocity in the horizontal circulation rolls reaches its maximum. As can be seen in Fig.2, apart from nearly axisymmetric wind field associated with the mesoscale Crimean cyclone (with orbital velocity of 4 − 5 $m \cdot s^{-1}$) there are also long and narrow circulation rolls; the distance between them is 3 − 5 km. As can be seen in an enlarged part of Fig.2, the rolls manifest themselves not only in the vertical velocity field but in the vertical vorticity field as well. Circulation rolls can also be observed outside the vortex area, along with circulation cells.

Fig.3 shows a vertical cross-section oriented normal to the rolls (see Fig.2 for orientation). As can be seen in Fig.3, the roll parameters are as follows: aspect ratio is 5:1, vertical velocity is up to ±0.4 − 0.6 $m \cdot s^{-1}$, horizontal velocity is ~ 0.3 $m \cdot s^{-1}$. The aspect ratio of 5:1 falls within the range of roll aspect ratio values determined from observation and numerical modeling [8].

It was mentioned earlier that the rolls are narrow with width of 3 − 5 km. So it is necessary to prove that 1-km resolution is sufficient for their modeling. In order to validate the main modeling results we used an additional nested domain with resolution of 300 m. Due to computational limitations the forth domain was relatively small and only the first stage of the Crimean vortex evolution could be modeled. Nevertheless, it was found that the circulation rolls in the 300-m resolution domain changed only insignificantly compared to the circulation rolls in the 1-km resolution domain.

The main mechanism of roll formation is convection in the PBL over the sea. Convective instability is characterized by the Rayleigh number that represents the ratio of buoyancy force to friction force $Ra = \frac{\Delta\theta \cdot g \cdot H^3}{\theta \cdot K_z \cdot K_T}$ where $\Delta\theta$ is a potential temperature drop in the PBL, $K_T$ is a heat diffusion coefficient [8]. In our case $\Delta\theta$ is 0.1 – 0.3 K, H is ~ 1 km, $K_z$ and $K_T$ are ~ 1 $m^2 \cdot s^{-1}$. Thus, we get Ra is ~ $10^6$. So, in our case the Rayleigh number exceeds the critical value, which means the onset of convection.

The ratio $z_i/L$ is also used as a criterion for the onset of convection. Here $z_i$ is the inversion height, L is the Monin-Obukhov length. The ratio $z_i/L$ allows estimating contribution from buoyancy force to generation of turbulent kinetic energy in an unstably stratified flow. The ratio $z_i/L$ for the vortex area (Fig.2) is −20...−16. According to [5] when $z_i/L > -5$ convection takes the form of rolls and when $z_i/L < -25$ convection takes the form of cells. So, the ratio of −20...−16 corresponds to mixed convection, which takes the form of both rolls and cells.

Fig.4 shows a vertical profile of potential temperature averaged over the occupied by the circulation rolls. As can be seen in Fig.4, the inversion height is ~ 800 m, and the rolls form in the lower almost neutrally stratified layer with thickness of 800 m. Sensible heat flux from the surface in the area occupied by the rolls is not large with values of 6 – 9 $W \cdot m^{-2}$.

So, indeed convective instability played an important role in the roll formation. It is well known that rolls caused by Rayleigh-Bernard convective instability are almost parallel to the background wind direction [8]. As can be seen in Fig.2, the rolls are not everywhere parallel to the wind direction in the vortex area. So, in our case there are other mechanisms of roll formation apart from convective instability.

Dynamic instability is also considered as one of the main mechanisms of roll formation [5]. A necessary condition for dynamic instability is that the wind velocity profile has an inflection point. However in our case the necessary condition is not satisfied (not shown), so dynamic instability can not occur.

Now, let us consider advection and stretching of vertical velocity field inhomogeneities (i.e. small areas of strong updrafts or downdrafts) as possible mechanisms of roll formation. Deformation rate can be estimated using the Okubo-Weiss criterion (OWC), which is defined as relative vorticity squared minus strain rate squared

$$W = (\sigma^2 - \zeta^2)/4$$

where $\sigma^2 = (u_x - v_y)^2 + (v_x + u_y)^2$, $\zeta^2 = (v_x - u_y)^2$, and $u_x$, $u_y$, $v_x$, $v_y$ are x- and y-derivatives of horizontal wind velocity components. It is well known that the OWC allows to determine a rate of particle scattering, i.e. rate of divergence of initially close trajectories [9, 10]. Depending on the sign of an OWC value there are two types of fluid motion: elliptic and hyperbolic. If OWC < 0 distance between Lagrangian particles changes periodically with time. If OWC > 0 distance between particles changes exponentially with time $\sim \exp\left(\pm\sqrt{W}t\right)$. So, the OWC allows comparing vorticity and deformation at a given point of a stationary flow [9, 10]. In our case using the OWC we can investigate how horizontal wind field influence the vertical velocity field inhomogeneities. One-sided stretching of such inhomogeneities leads to appearing of long and narrow "bands" in the vertical velocity field.

In order to separate the above mentioned mechanism of roll formation, two numerical experiments were performed. In the numerical experiments convective instability is artificially suppressed but advection and deformation of the vertical velocity field inhomogeneities is preserved.

**Circulation rolls in the numerical experiments**

In the first experiment sea surface temperature (SST) in the initial conditions is decreased by 2 K and in the second experiment SST is decreased by 10 K. All other fields in the initial conditions are the same as in the control run. Presumably the SST decrease suppresses convective instability and allows detecting other mechanisms of roll formation.

As a result of the SST decrease the Crimean cyclone lifetime shortened by 20 % in the first experiment (with 5 K-SST decrease) and by 30 % in and in the second experiment (with 10 K-SST decrease). However the vortex structure did not change significantly. Its height, radius, and maximum orbital velocity also remained almost unchanged in both experiments. As expected, the PBL height drastically decreased. Fig.5 shows potential temperature profile for the second experiment (with 10 K-SST decrease). In the control run the PBL height is 800-1000 m, but in the numerical experiment the PBL is just a thin surface layer with thickness of 50-100 m and there is a stably stratified layer with thickness of ~ 400 m above the PBL. Also, there is no sensible heat flux from the surface. It is evident that convection can not develop in such PBL. The necessary condition for dynamic instability is satisfied neither in the numerical experiments nor in the control run.

Fig.6 shows vertical and horizontal wind fields from the second numerical experiment (with 10 K-SST decrease). As can be seen in Fg.6 there are long “bands” spiraling to the vortex center that resemble the circulation rolls in the control run (Fig.2). Besides, there are short rolls in the eastern part of the vortex.

Let us consider two possible mechanisms of roll formation. Foremost, the roll formation can be explained by positive values of the OWC field. Fig.7 shows OWC field for the Crimean vortex spatially smoothed over 121 computational grid points and averaged over 2 hours from 0600 LT to 0700 LT 16 August. The vortex began to move away from the shore at 0400 LT, so the rolls in Fig.7 developed during ~2 − 3 hours while the vortex was moving over the sea. As can be seen in Fig.7, the OWC values are large and positive in the eastern part of the vortex (~ $10^{-7}$ $s^{-2}$), which can explain the development of short circulation rolls there (Fig.6). During t = 2 − 3 hours distance between initially close particles increases by $\exp\left(\sqrt{10^{-7}} \cdot t\right) \approx 10 - 30$ times.

Also, the rolls can result from advection of vertical velocity field local inhomogeneities. As can be seen in Fig.6 there are several long rolls that spiral from the coastal area to the vortex center. They result from advection of vertical velocity field local inhomogeneities that always appear when an air flow comes to the shore. Indeed, in our case wind speed is 5 − 10 m·s$^{-1}$. It means that during t = 2 − 3 hours coastal inhomogeneities are advected over 30 − 100 km, which corresponds to the modeling results.

It is of interest to note that widening of the inhomogeneities caused by horizontal diffusion is negligible. Indeed, horizontal diffusion rate is equal to $\sqrt{K_h / t}$ ~ 200 m·h$^{-1}$. So the lateral widening of the rolls during advection is a small value of 400 − 600 m Also it seems that horizontal wind field itself prevents the inhomogeneities from widening (Fig.6). As can be seen in Fig.7, W>0 in the area occupied by the roll-like “bands”, i.e. the vertical wind field inhomogeneities can only shrink during advection.

Fig.8 shows a vertical cross-section of a long roll depicted in Fig.6. As can be seen in Fig.8, the horizontal roll is small and low. Its height is ~ 200 m, width is ~ 3 km, and circulation velocity is ~ 0.05 m·s$^{-1}$. It should be noted that the roll structure does not change significantly when the domain spatial resolution is increased to 300 x 300 m.

The coastal vertical wind field inhomogeneities that are advected from the coast to the open sea do not always turn into the long rolls. Sometimes there is a narrow area of strong air ascending in the almost vertically homogeneous PBL and a vast area of weak air subsidence (left part of Fig.2 and Fig.6).

The “bands” in the numerical experiments are rather long lived with lifetime of several hours. Essentially, atmospheric phenomenon lifetime is governed not by Newtonian friction law but by linear friction in the Ekman boundary layer. This is true in our case as well. The Crimean cyclonic vortex lifetime was determined by its height and friction in the Ekman boundary layer [2]. The Crimean vortex had lifetime of 12 − 15 h and height of 1.5 − 2 km. The vertical velocity field inhomogeneities are lower with height of 150 − 200 m, so their lifetime is about 3 − 4 hours. Nevertheless, it was enough time for the circulation rolls to form. Anyway, these rolls are of indisputable interest and their study will be the object of another paper.

**Conclusion**

In this paper we study circulation rolls that appeared in atmospheric modeling results for the Black Sea Region. Convective and dynamic instability as well as advection and stretching of vertical velocity field inhomogeneities are considered as possible mechanisms of roll formation. It is shown that convective instability indeed played an important role in the roll formation. Dynamic instability, which is considered as another important mechanism of roll formation, developed neither in the control run nor in the numerical experiments. In the first and second numerical experiments SST in the initial conditions was decreased by 5K and 10K respectively in order to suppress convective instability. However roll-like “bands” still appeared in the modeling results, although they were weaker than the circulation rolls from the control run. It is shown that the “bands” were caused by advection and stretching of vertical velocity field inhomogeneities.

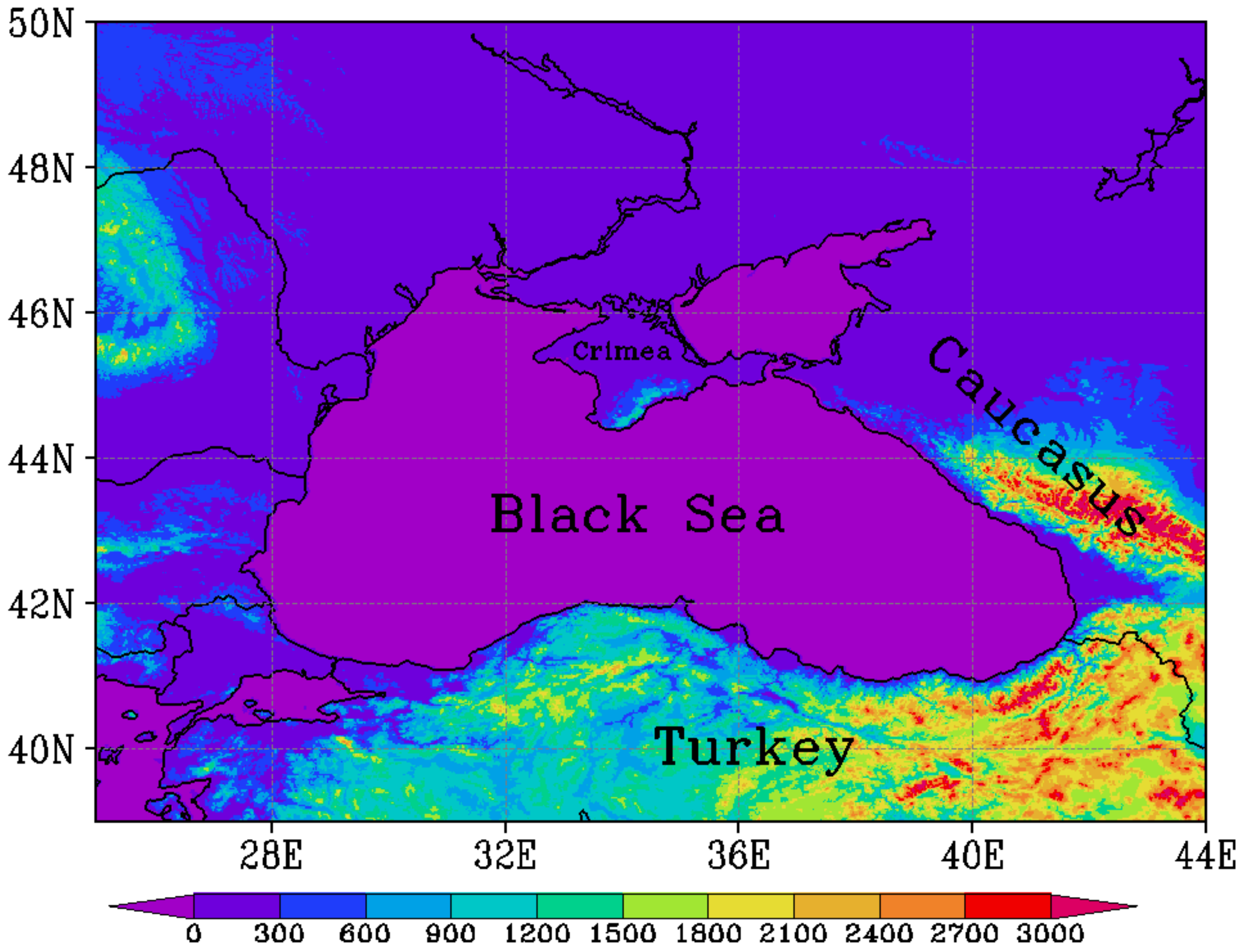

Fig. 1 A map of the geographical regions that are considered in this paper

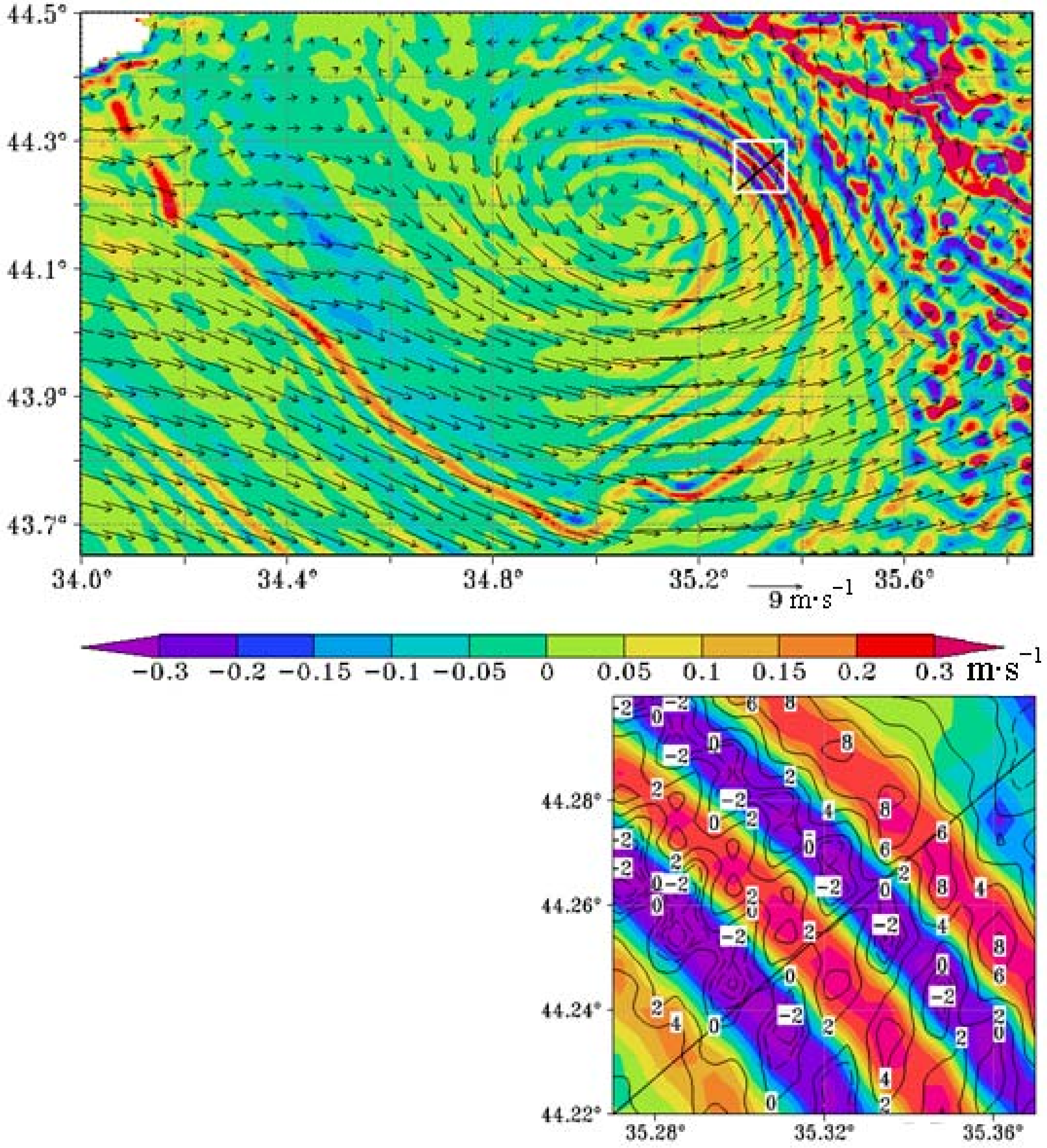

Fig.2 Vertical velocity field (m·s$^{-1}$, color) and horizontal wind field (m·s$^{-1}$, arrows) at height of 400 m at 0900 LT 16 August 2007 for the control run. An enlarged part of Fig.2 shows vertical velocity field (color) and vorticity ($10^{-4}$ c$^{-1}$) field (isolines)

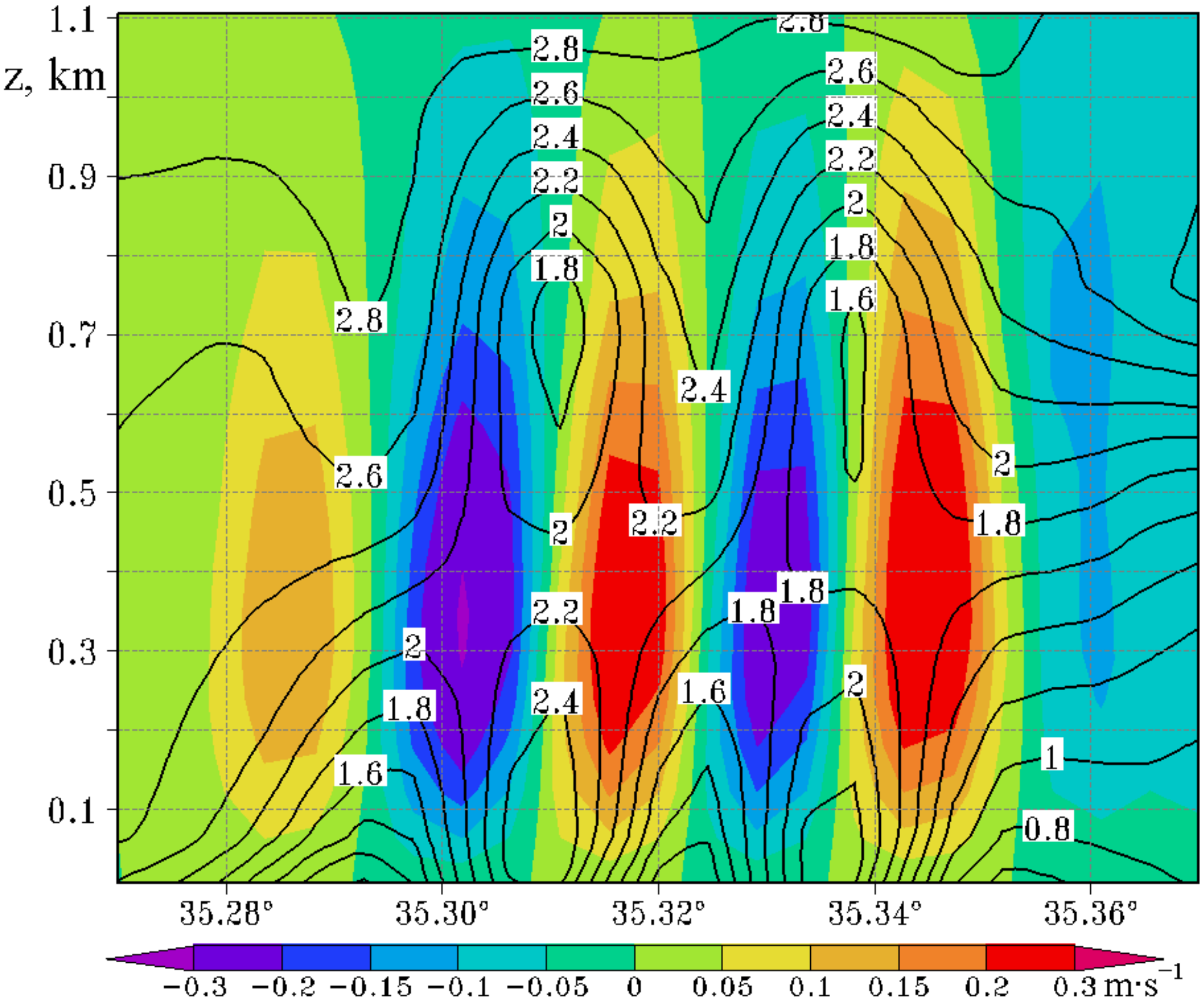

Fig.3 Vertical cross-section (see Fig. 2 for orientation) of vertical velocity (color) and horizontal velocity parallel to the cross section (isolines)

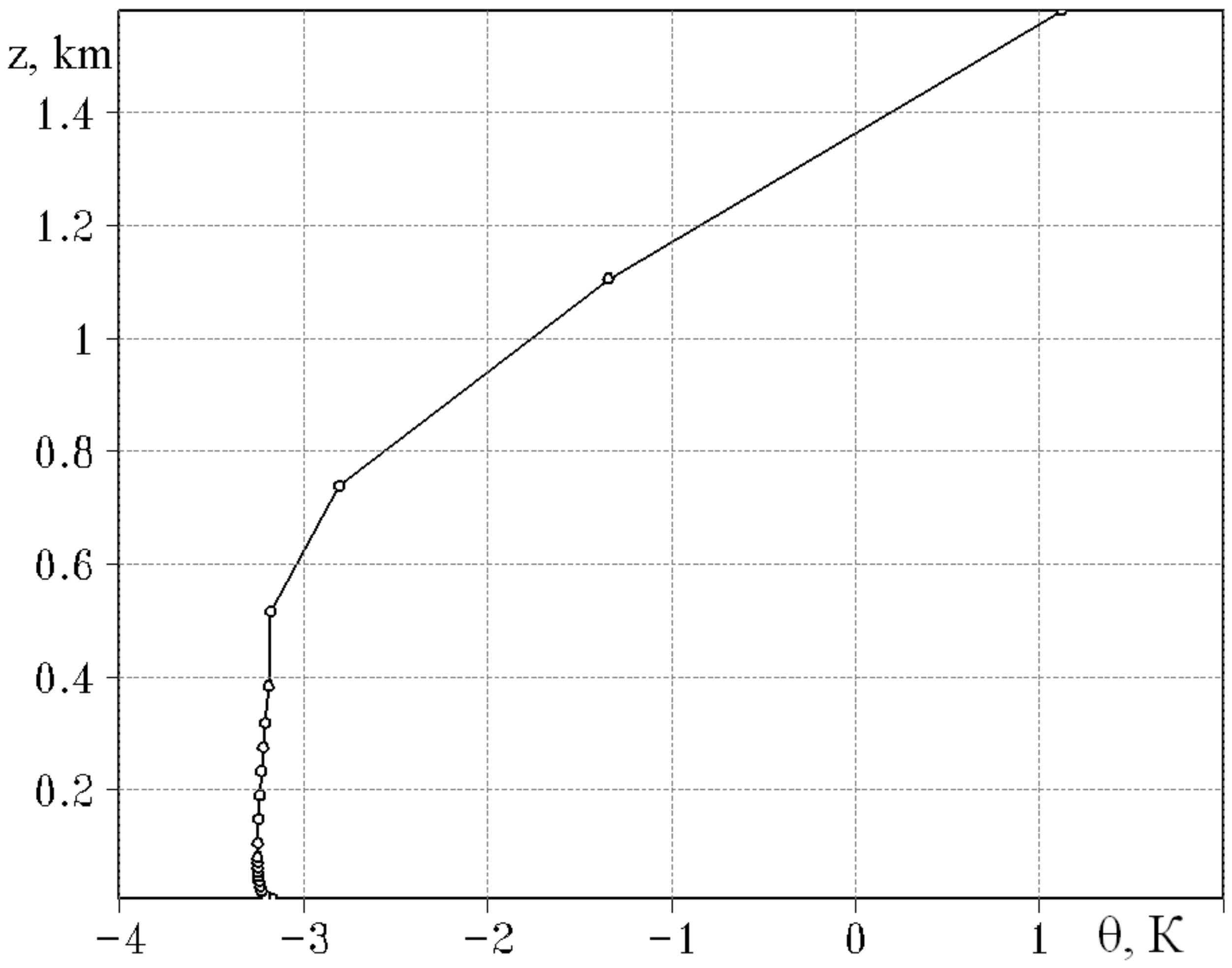

Fig.4 Potential temperature (K) profile for the control run. Potential temperature is spatially averaged over the rectangular area shown in Fig.2 at 0900 LT 16 August 2007

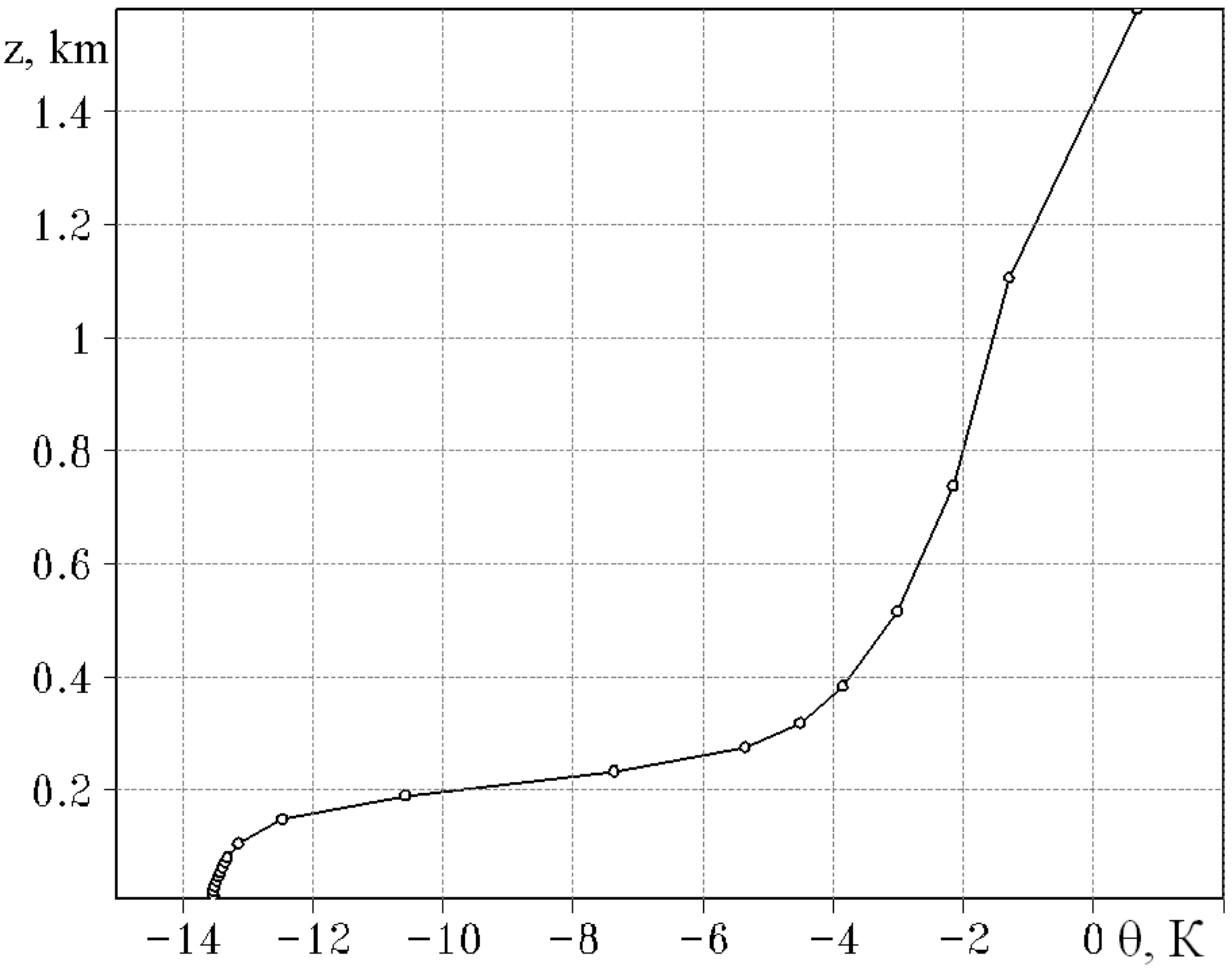

Fig.5 Potential temperature (K) profile for the experiment with 10 K-SST decrease. Potential temperature is spatially averaged over the rectangular area shown in Fig.6 at 0700 LT 16 August 2007

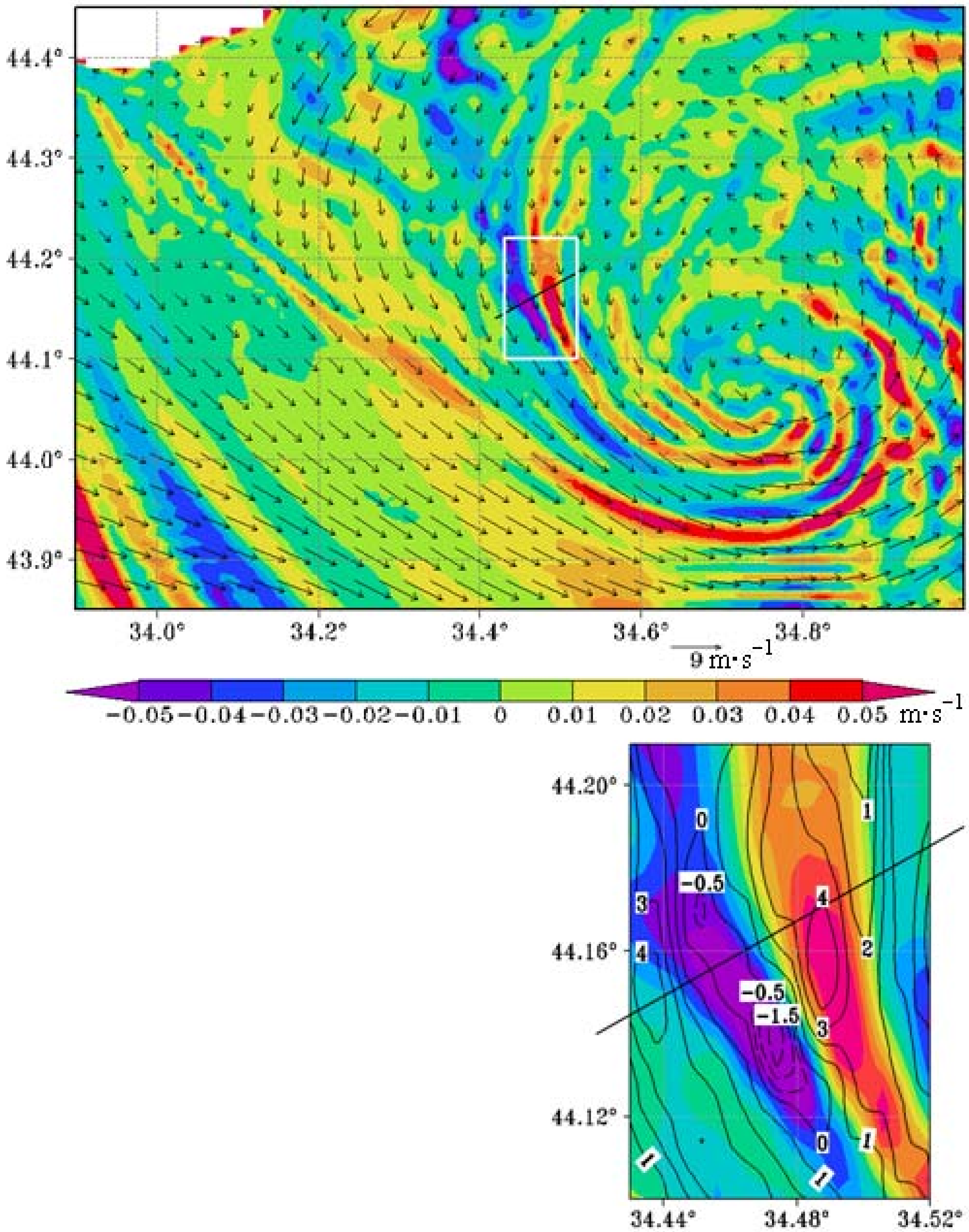

Fig.6 Vertical velocity (color) and horizontal wind field (arrows) at height of 100 m at 0700 LT 16 August 2007 for Experiment with 10 K-SST decrease. An enlarged part of Fig.6 shows vertical velocity field (color) and vorticity ($10^{-4}$ $s^{-1}$) field (isolines)

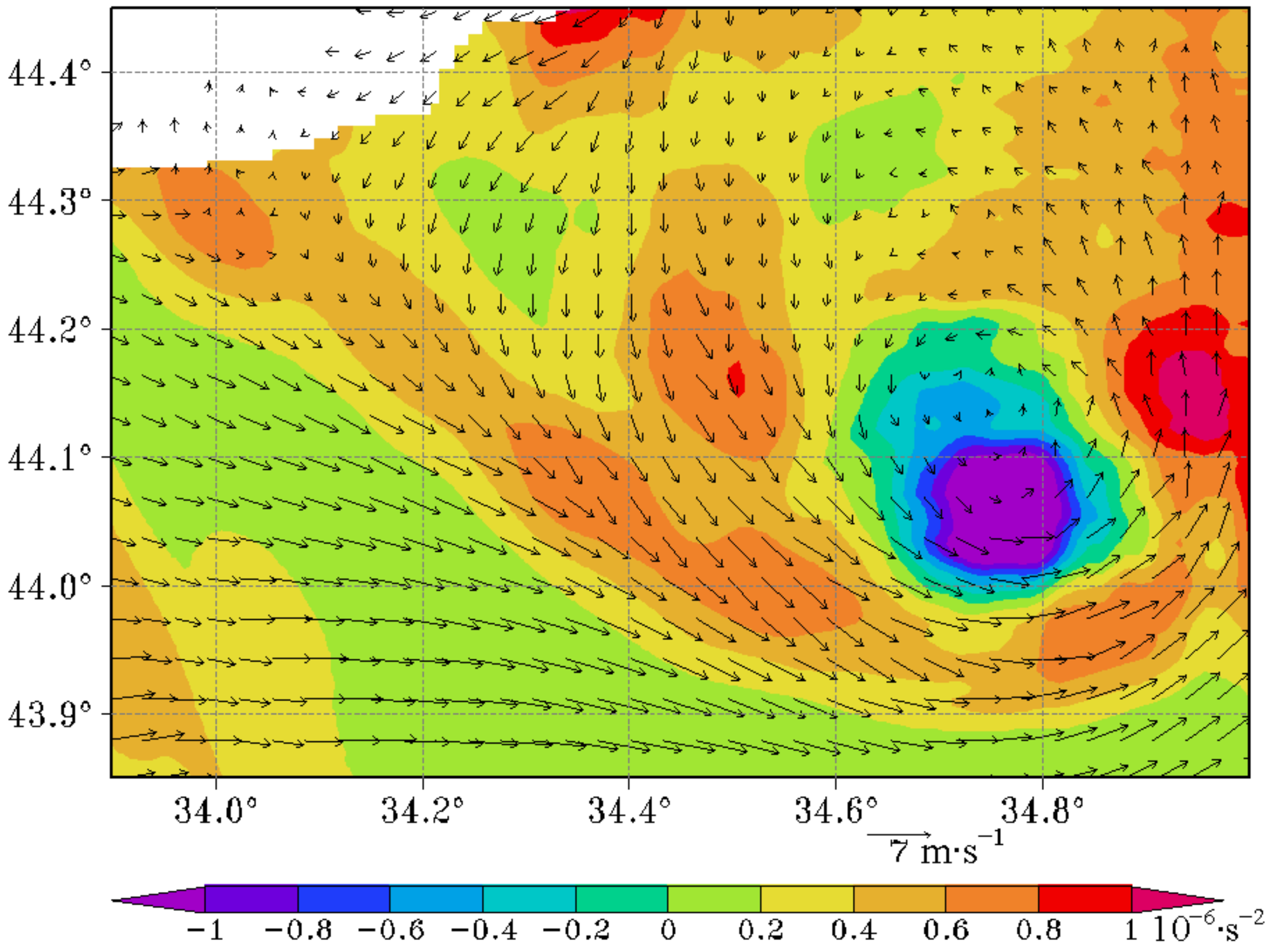

Fig.7 OWC field ($10^{-6}$ $s^{-2}$) and surface horizontal wind field (arrows) for the experiment with 10 K-SST decrease. OWC field is spatially smoothed over 121 points and averaged over 2 hours from 0600 LT to 0700 LT 16 August 2007

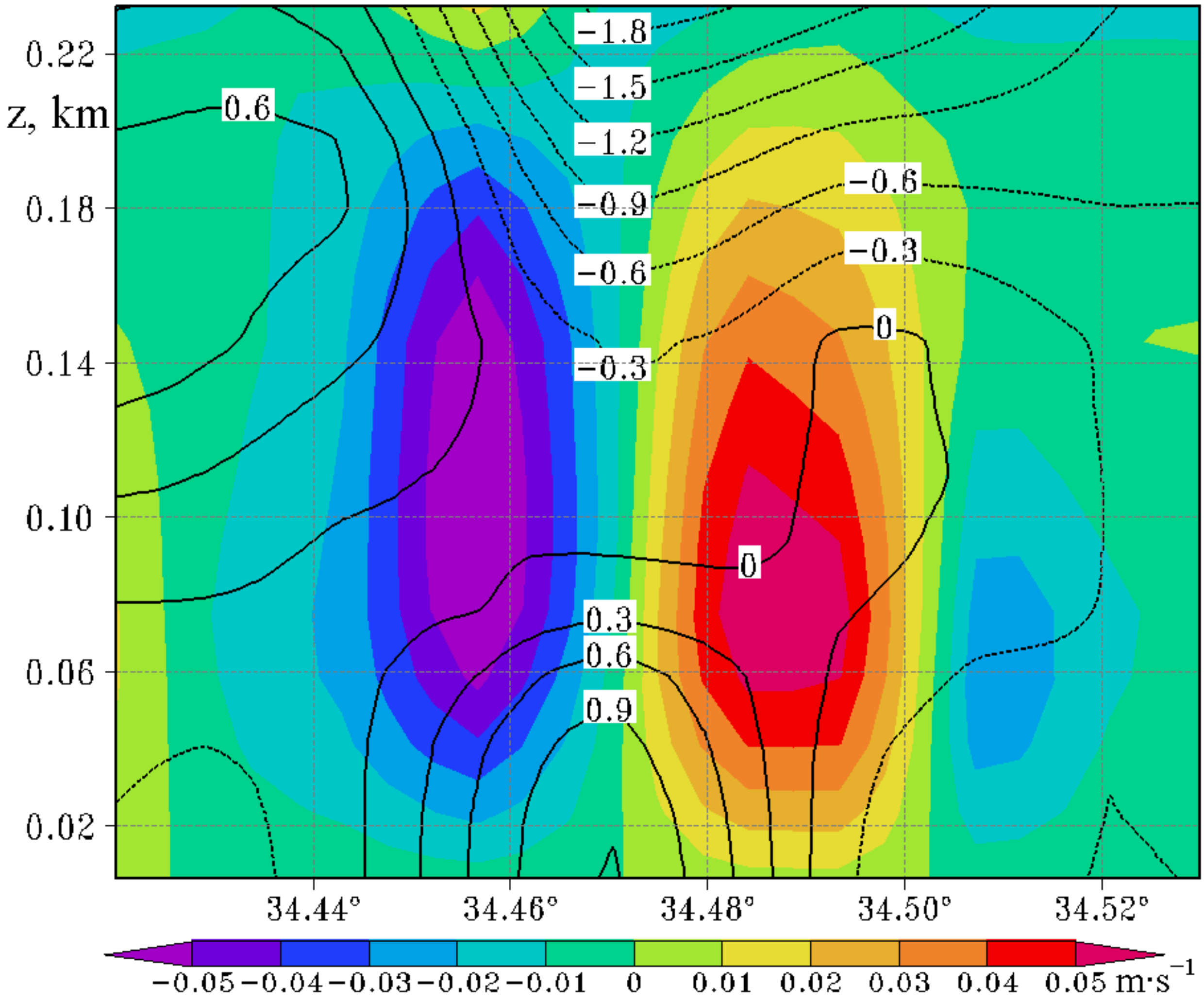

Fig.8 Vertical cross-section (see Fig.6 for orientation) of vertical wind speed ($m \cdot s^{-1}$, color) and horizontal velocity parallel to the cross-section (isolines)